\documentclass[sigplan]{acmart}

\setcopyright{none}

\usepackage{booktabs}   
\usepackage{subcaption} 
\usepackage[linesnumbered,ruled,lined]{algorithm2e}               
\usepackage{epstopdf}
\usepackage{multirow}
\usepackage{listings}
\newcommand{\ie}{{\it i.e.,}}
\newcommand{\eg}{{\it e.g.,}}

\begin{document}
	
	\title{Machine Learning Framwork for Performance Anomaly in OpenMP Multi-Threaded Systems}         

	
	\author{Weidong Wang}
	
	\orcid{0000-0002-7378-2766}             
	\affiliation{
		\department{Faculty of Information Technology}              
		\institution{Beijing University of Technology}            
		\streetaddress{Chaoyang District Pingleyuan No.100}
		\city{Beijing}
		\state{}
		\postcode{100124}
		\country{China}                    
	}
	\email{wangweidong@bjut.edu.cn}          
	
	\author{Wangda Luo}
	
	\orcid{nnnn-nnnn-nnnn-nnnn}             
	\affiliation{
		\department{Faculty of Information Technology}             
		\institution{Beijing University of Technology}           
		\streetaddress{}
		\city{Beijing}
		\state{}
		\postcode{}
		\country{China}                   
	}
	\email{luowangda_bjut@163.com}         

\begin{abstract}
Some OpenMP multi-threaded applications increasingly suffer from performance anomaly owning to shared resource contention as well as software- and hardware-related problems. Such performance anomaly can result in failure and inefficiencies, and are among the main challenges in system resiliency. To minimize the impact of performance anomaly, one must quickly and accurately detect and diagnose the performance anomalies that cause the failures. However, it is difficult to identify anomalies in the dynamic and noisy data collected by OpenMP multi-threaded monitoring infrastructures. This paper presents a novel machine learning framework for performance anomaly in OpenMP multi-threaded systems. To evaluate our framework, the NAS Parallel NPB benchmark, EPCC OpenMP micro-benchmark suite, and Jacobi benchmark are used to test the performance of our framework proposed. The experimental results demonstrate that our framework successfully identifies 90.3\% of injected anomalies of OpenMP multi-threaded applications.

\end{abstract}



\maketitle

\section{Introduction}

Extreme-scale computing is expected to involve hundreds of millions of processes and/or threads with multi-level parallelism running on large-scale hierarchical and heterogeneous hardware. Such processes and/or threads become so important because they are directly or indirectly related with performance variation, which are derived from nearly hardware and software associated with anomalies such as orphan processes and/or threads left over from previous jobs \cite{Radojkovi2016}, memory leak, and system slow \cite{Vargas2017,Bhatele2013}. In addition to performance degradation, these anomalies can also lead to premature OpenMP job terminations \cite{Agullo2017}. The unpredictability caused by anomalies, combined with the growing size and complexity of OpenMP applications, makes efficient management challenging, becoming one of the roadblocks on the design of extreme-scale parallel computing \cite{Seo2018}.

Detection and diagnosis of anomalies, \eg\ the correctness, failure, and resilience of runtime programs, have heavily relied on more expertise-based judgments. On one hand, by continuously monitoring and analyzing system logs and application resource usage patterns, OpenMP operators \cite{Ayguade2009} can assess program health and identify the root causes of anomalies. Actually, in most of OpenMP applications, this process would challenge to translate into the manual analysis of thousands of data points per day [10]. As the size of OpenMP applications grows, such manual processing becomes increasingly time-consuming and error-prone \cite{Ibidunmoye2015}. Hence, automated anomaly diagnosis is crucial for the efficient operation of future OpenMP applications.

In the traditional field of computer research, such as underlying architecture \cite{Marongiu2012,Yamazaki2018} and multiple threads \cite{Aldea2016} based software design, the study heartbeats in a multi-threaded program has gradually become a hot spot in both industry and academia. According to the analysis of heart rate, we can better understand the information of execution state and exception, and provide a guarantee for program running reliability. While many techniques have been proposed for detecting the root cause of anomalies \cite{Hiranya2017} in OpenMP applications, these techniques still rely on human directors to identify the root causes of the anomalies, leading to wasted computing resources. An effective way of decreasing the impact of anomalies is to automate the diagnosis of anomalies \cite{Yu2016}, which lays the foundation for automated diagnosis.

In this paper, we propose a framework to automatically diagnose OpenMP applications suffering from previously observed anomalies at runtime and classify the possible anomaly states of these applications. Comprehensive experiments are conducted to study accuracy and F-Score of our proposed framework. The experimental results show the high accuracy of our framework.

\begin{figure*}[t]
	\centering
	\includegraphics[width=16cm]{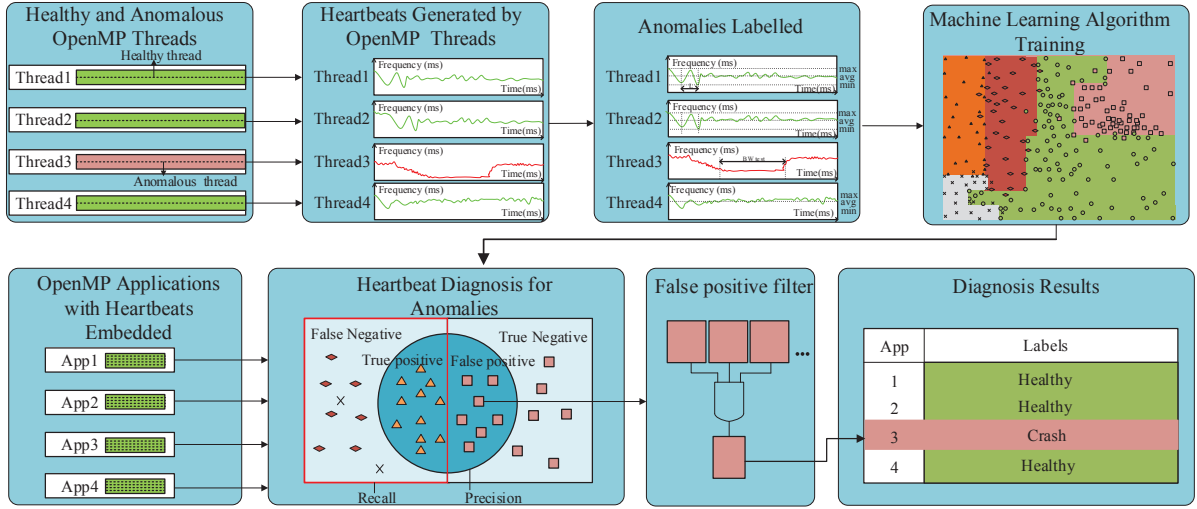}
	\caption{Overall system architecture.}\label{fig:ora}
\end{figure*}

\section{Machine Learning Framework}
\label{sec:roadmap}

Our goal is to accurately detect whether an OpenMP thread is anomalous (\ie\ experiencing anomalous behavior) and classify the type of the anomaly (e.g., shutdown or memory leak) at runtime. We target anomalies that are caused by applications or system software/hardware such as anomaly termination and memory leak. To learn most behaviors of OpenMP applications at runtime, based on the heartbeat APIs that can be injected into these OpenMP programs, they will generate regular heartbeat sequences while the OpenMP programs run. To diagnose anomalies, we implement a novel machine learning framework via a heartbeat diagnosis of performance anomaly for OpenMP multi-threaded applications.

Figure \ref{fig:ora} shows an overview of our framework. We leverage historical heartbeat data that are collected from healthy and anomalous OpenMP applications to learn the behavior of target anomalies. As an OpenMP application runs on multiple threads in parallel, if any of these threads is anomalous, its entire threads heartbeat patterns may be affected. Hence, if a job running on multiple threads suffers from an anomaly, we include only the anomalous thread in the training set and discard the remaining threads' data.

Through the heartbeat data collected from known healthy and anomalous runs, we identify the statistical features that are useful to detect target anomalies, and analyze concise anomaly features using machine learning algorithms. 

\subsection{Heartbeat Data Collector}
\label{subsec:AALH}

To diagnose OpenMP applications' health, we collect the heartbeat data generated from the individual OpenMP application that is used in the known healthy and anomalous runs. we leverage heartbeat data that are already periodically collected from each OpenMP working thread among these OpenMP applications. These heartbeat data typically consist of the thread ID, timestamp, and heart rate. It is well-known that most anomalies commonly show some portents. These portents, also called features, can be collected by experiences. So we first manually produce all possible anomalies, and inject them into these OpenMP applications. Then, we record the heartbeat data from all the abnormal working treads.

Moreover, we use these heartbeat time series to train supervised machine learning models where the label of each OpenMP applications is given as the type of the observed anomaly on that application (or healthy). In the absence of labeled data for anomalies, the training can be conducted using experiments with synthetic anomalies, which are programs embedded by simulating real-life anomalies. With training data from a diverse set of OpenMP applications that represent the expected anomalous or healthy runs, the machine learning algorithms extract the characteristics of anomalies independent of the OpenMP applications. This allows us to identify previously observed anomaly characteristics among heartbeat time series from different OpenMP applications.

\subsection{Feature Extraction}
\label{subsec:FE}

To judge whether a heartbeat sequence differs from a normal sequence, we define the basic features of a heartbeat sequence based on heart rate, completion time, lower bound, and similarity.

Suppose there are two sequences $Q$ and $C$ as shown in Equation \ref{formu11}. And the length is $m$ and $n$, respectively.

\begin{equation}
	\footnotesize
	\begin{aligned}
		& Q={(t_1,q_1),(t_2,q_2),(t_3,q_3),\cdots,(t_n,q_n)} \\
		& C={(t'_1,c_1),(t'_2,c_2),(t'_3,c_3),\cdots,(t'_m,c_m)}
	\end{aligned}
	\label{formu11}
\end{equation}

To measure the ratio of completion time, the global time ratio of two heartbeat sequences is defined in Equation \ref{formu132}.

\begin{equation}
	\footnotesize
	GlobalTimeRatio(C,Q)=\frac{t'_m}{t_n},
	\label{formu132}
\end{equation}

\noindent where $t'_m$ is the completion time of heartbeat sequence $C$, and $t_n$ is the completion time of heartbeat sequence $Q$. To measure the local differences of completion time, we employ a local fine-grained time ratio based on the sliding window as follows.

\begin{equation}
	\footnotesize
	LocalTimeRatio(C,Q,w,k)=\frac{\sum\limits_{i=1}^n \frac{t'_{i+w}-t'_{i}}{t_{i+w}-t_{i}}}{k},
\end{equation}

\noindent where $t'_{i+w}$ denotes the completion time of sequence $C$ at the ($i+w$)-th timestamp. And $t_{i+w}$ denotes the completion time of sequence $Q$ at the ($i+w$)-th timestamp. $w$ is the size of sliding window. $k$ represents the number of sliding windows.

To measure the ratio of heart rate between two heartbeat sequences, in this paper the heartbeat ratio is defined in Equation (\ref{formu133}).

\begin{equation}
	\footnotesize
	GlobalHeartbeatRatio(C,Q)=\frac{\frac{\sum\limits_{i=1}^mc_i}{m}}{\frac{\sum\limits_{i=1}^nq_i}{n}},
	\label{formu133}
\end{equation}

\noindent where $m$ and $n$ represent the length of heartbeat sequence $C$ and $Q$. $c_i$ and $q_i$ are the heart rate of $i$-th heartbeat belong to sequence $C$ and $Q$. To measure the local differences of heartbeat ratio, we employ a local fine-grained heartbeat ratio based on the sliding window as follows.

\begin{equation}
	\footnotesize
	LocalHeatbeatRatio(C,Q,w,k)=\frac{\sum\limits_{i=1}^n \frac{c_{i+w}-c_{i}}{q_{i+w}-q_{i}}}{k},
	\label{formu1355}
\end{equation}

\noindent where $c_{i+r}$ denotes the heart rate of sequence $C$ at the ($i+w$)-th timestamp. And $t_{i+w}$ denotes the heart rate of sequence $Q$ at the ($i+w$)-th timestamp. $w$ represents the number of sliding windows. $k$ is the size of sliding window.

To measure a similarity between two heartbeat sequences, DTW, dynamic time wrapping \cite{ding2013developments} is employed for sequence analysis. The DTW distance can be found by an optimal bending path to minimize the cumulative distance of two heartbeat sequences. 

The DTW distance is calculated between two heartbeat sequences as follows.

\begin{equation}
	\footnotesize
	DTW(q_i,c_j)=\sum\limits_{i=0}^m \sum\limits_{j=0}^n D(q_i,c_j),
	\label{formu6}
\end{equation}

\noindent where $D(q_i,c_j)$ is the dynamic time warping function, which can obtain an optimal match between two given heartbeat sequences with certain restrictions and rules as follows.

\begin{equation}
	\footnotesize
	\left\{
	\begin{aligned}
		&0, (i=0,j=0) \\
		&\left|q_i-c_j\right|+D(q_{i-1},c_j),(i\geq 1,j=0) \\
		&\left|q_i-c_j\right|+D(q_{i},c_{j-1}), (i=0,j\geq 1)\\
		&\left|q_i-c_j\right|+\min[D(q_{i-1},c_{j-1}),D(q_{i-1},c_j),D(q_i,c_{j-1})], i,j \geq 1\\
	\end{aligned}
	\right.
	\label{formu6-7}
\end{equation}

\noindent where $|q_i-c_j|$ represents Euclidean distance between the heart rate of $i$-th heartbeat in the sequence $C$ and the heart rate of $j$-th heartbeat in the sequence $Q$. The optimal match is denoted by the match that satisfies all the restrictions above and that has the minimal cost, which is calculated as the sum of absolute differences for each matched pair of indices between their values. Hence, the minimal cost distance can be easily calculated by Dynamic Programming \cite{bertsekas1995dynamic}. For the sake of processing speed, as the complement of DTW distance, we also introduce the LB\_Keogh lower bound \cite{keogh2006lb_keogh} since it can filter most of the sequences that cannot be the optimal matching heartbeat sequences as follows.

First, we define upper bound sequence $u=u_1,u_2,\cdots,u_n$ for the heartbeat sequence $Q=q_1,q_2,\cdots\,q_n$ as follows.

\begin{equation}
	\footnotesize
	\left\{
	\begin{aligned}
		&u_i = \max(q_0,q_1,\cdots,q_{i+w}), i<w \\
		&u_i = \max(q_{i-w},q_{i-w+1},\cdots,q_{i+w}), i\geq w \\
	\end{aligned}
	\right.
	\label{formu10}
\end{equation}

Similarly, we also define lower bound sequence $l=l_1,l_2,\cdots,l_n$ for the heartbeat sequence $Q=q_1,q_2,\cdots\,q_n$ as follows.

\begin{equation}
	\footnotesize
	\left\{
	\begin{aligned}
		&l_i = \min(q_0,q_1,\cdots,q_{i+w}), i<w \\
		&l_i = \min(q_{i-w},q_{i-w+1},\cdots,q_{i+w}), i\geq w \\
	\end{aligned}
	\right.
	\label{formu101}
\end{equation}

\begin{table*}[t]
	\center
	\caption{Performance comparisons. LR denotes logistic regression. DT denotes decision tree. RF denotes random forest. NB denotes Naive Bayes. HSA denotes the heartbeat sequence analysis method. N represents normal status. A represents abnormal status. S represents shutdown status.}
	\begin{tabular}{lrrcccccccccccccccc}
		\toprule
		\multirow{3}*{Method} &  \multicolumn{3}{c}{NPB-sp}  &  \multicolumn{3}{c}{NPB-lu} &  \multicolumn{3}{c}{NPB-bt} &  \multicolumn{3}{c}{NPB-cg}   & \multicolumn{3}{c}{EPCC-Array} &  \multicolumn{3}{c}{Jacobi}\\
		\cmidrule(r){2-4}  \cmidrule(r){5-7} \cmidrule(r){8-10}  \cmidrule(r){11-13}  \cmidrule(r){14-16}  \cmidrule(r){17-19}
		& N &A &S & N &A &S & N &A &S        & N &A &S & N &A &S & N &A &S \\
		\midrule
		LR & 0.55 &0.87 &0.91        & 0.63 &0.88 &0.58              & 0.44  &0 &0            & 0.65 &0 &0.84              & 0.64  &0.90  &0              & 0.44 &0 &0       \\
		
		DT       & 0.86 &0.91 &0.92        & 0.42 &0.79 &0.62              & 0.45 &0.70 &0.55       & 0.94 &0.94 &1.00           & 1.00  &0.31  &0.83           & 0.63 &0.07 &0.31        \\
		
		RF       & 0.67 &0.81 &0.63        & 0.46 &0.93 &0.53              & 0.43 &0.66 &0.40       & 0.94 &0.93 &1.00           & 1.00  &0.93  &0.95           & 0.62 &0.14 &0.32         \\
		
		NB         & 0.49 &0.88 &0           & 0.52 &0.71 &0.08              & 0.37 &0 &0             & 0.50 &0    &0              & 0.53  &0     &0              & 0.44 &0 &0         \\
		
		SGDC                & 0    &0.81 &0.87        & 0.53 &0.73 &0.12              & 0.46 &0 &0.12          & 0.86 &0    &0.74           & 0     &0.68  &0.48           & 0.48 &0.28 &0.47         \\
		
		HSA                 & 0.93 &0.89 &0.91        & 0.86 &0.90 &0.85              & 0.83 &0.78 &0.90       & 1.00  &0.86    &1.00            & 1.00  &0.90  &1.00           & 0.96 &0.75 &1.00\\
		\cline{1-19}
		\bottomrule
	\end{tabular}
	\label{tab4-3}
\end{table*}

\noindent where $w$ denotes the size of the sliding window. Hence, for the given heartbeat sequence $Q$, we can obtain two sequences, \ie\ upper bound heartbeat sequence $u$ and lower bound sequence $l$.

Then, we calculate the LB\_Keogh value between the given heartbeat sequence $Q$ and the target heartbeat sequence $C$ in Equation \ref{formu10}.

\begin{equation}
	\footnotesize
	LB\_Keogh(Q,C)=\sum\limits_{i=1}^n \left\{
	\begin{aligned}
		&(c_i-u_i)^2,c_i>u_i \\
		&(c_i-l_i)^2,c_i<u_i\\
		&0, otherwise
	\end{aligned}
	\right.
	\label{formu10}
\end{equation}

\noindent where $c_i$ denotes $i$-th heartbeat data in the target heartbeat sequence $C=c_1,c_2,\cdots,c_n$. The accumulative differences are calculated by Euler distance among the upper bound $u$ and lower bound sequence $l$ of the given heartbeat sequence $Q$.

\subsection{Training}
\label{sec:experiments}

To train the efficacy of our framework, we run controlled experiments on a multi-core computer environment. We mimic anomalies observed in the environment by running synthetic programs simultaneously with various OpenMP applications, and diagnose the anomalies using our framework and selected benchmarks. For the purpose of experimental reproducibility, we performed our experiments at an acceptable cost using a public cloud resource. Meanwhile, we built the heartbeat dataset based on famous OpenMP benchmarks. This section describes the details of the target experiments.

We employ the Tencent cloud multi-core computer with standard \emph{S2.LARGE8} v4 CPU, 8G memory, and 5Mbps as the running environment for OpenMP benchmark applications. Then, we run 64-bit Linux version 5.4.0 and GCC version 9.3.0 as the operating system and OpenMP compiler, respectively. The details of the heartbeat dataset based on OpenMP Benchmarks are shown in the following sections.

We implement our framework in \emph{C} for heartbeat generation and \emph{Python} for heartbeat diagnosis, respectively. First, we use \emph{C} standard library to implement heartbeat function API for OpenMP applications. Then, we use the \emph{pandas} and \emph{sklearn} package for machine learning algorithms including \emph{logistic regression},\emph{ decision tree}, \emph{random forest}, naive \emph{bayes}, and \emph{SGDC classifier} during performance comparisons. To enhance the accuracy of the analysis, we have two strategies.

\subsection{Testing and Results}
\label{sec:experiments}

To test our framework, we use a heartbeat dataset collected from representative OpenMP benchmark applications. In our testing, we use three different types of OpenMP benchmarks \textit{i.e.} NPB, EPCC, and Jacobi. (1) The NAS Parallel Benchmarks (NPB) are widely utilized by the parallel computing community as a representative set of OpenMP applications. (2) The EPCC OpenMP micro-benchmark suite is developed by Edinburgh Parallel Computing Centre for measuring the overheads of synchronization, loop scheduling, and array operations in the OpenMP runtime library. These benchmarks run in multiple cores of various scientific workloads. And (3) The Jacobi is another scientific computing application for multiple system performance analysis.

The heartbeat dataset provides a group of .xlsx files that record various heartbeats with the benchmark applications above. 

\subsubsection{Measures}

We employ the following statistical measures to assess how well the heartbeat diagnosis framework work in different anomalies:

\begin{itemize}
	\item Accuracy: The fraction of the number of correctly predicted to the number of all predictions.
	\item Precision: The fraction of the number of windows correctly predicted with an anomaly type to the number of all predictions with the same anomaly type.
	\item Recall: The fraction of the number of windows correctly predicted with an anomaly type to the number of windows with the same anomaly type.
	\item F-Score: The harmonic mean of precision and recall.
	\item Macro F-Score: The F-score calculated using the weighted averages of precision and recall, where the precision and recall of each class is weighted by the number of instances of that class.
\end{itemize}

\subsubsection{Results}
\label{sec:experiments}

In our experiments, we randomly select 30\% the same size of training samples in the dataset for each machine learning method while the remaining heartbeat data is used as testing samples.

Then, we use a local desktop computer \ie\ Intel i5-9400F CPU, 16G memory, and 512G HDD to execute the heartbeat diagnosis program. Then, we choose 64-bit Ubuntu version 18.04 and Python 3.6 as the operating system and coding language, respectively. 

To avoid the experimental deviation by randomly selecting samples, we repeat the process above three times for each experiment, and take the average of experimental results.

To study our heartbeat diagnosis framework, we employ a series of typical supervised methods of machine learning such as Logistic Regression, Decision Tree, Random Forest, Naive Bayes, and SGD classifier. To make sense of the process of anomaly diagnosis, we divided heartbeat samples into three groups, \ie\ normal group, memory leak group, and shutdown group. For each group, we test macro F-score using the above machine learning methods. As shown in Table \ref{tab4-3}, the macro F-score of the HSA outperforms other machine learning methods in the following OpenMP benchmarks, \ie\ NPB-sp, NPB-lu, NPB-bt, NPB-cg, EPCC-array, and Jacobi. For instance, in Benchmark NPB-cg, the macro average F-score is 0.95 that is beyond the decision tree 0.94 and random forest 0.94, although its F-score of the memory leak group is 0.86 that is below the decision tree 0.94 and random forest 0.93. The above results show that the entire our framework works well.

In addition, it can be observed from Table \ref{tab4-3} that there are some zero values that occurred on the other competitors. For instance, 0 is at Row 1 and Column 8 for NPB-bt, which shows that the LR method does not work in diagnosing memory leak anomaly for the NPB-bt dataset. This is because the LR method could not effectively learn such characteristics of heartbeat data in the limited number of training samples. 

\section{Conclusion}
\label{sec:con}

Complementary to previous performance anomaly diagnosis approaches, we implemented a novel machine learning framework.

In the work, we proposed the anomaly diagnosis framework of multi-threaded OpenMP applications that enables automatic detection and diagnosis of previously observed anomalies. Compared with traditional multi-threaded diagnosis technology, our customized framework typically considered multiple parameters from multiple-threaded OpenMP applications at runtime.

For the future work, we may consider several aspects to further improve the performance of the framework proposed. For further improving the accuracy of the HSA, we will cluster more feature metrics from different dimensions of OpenMP runtime into some representative ones using an improved k-means clustering method. Furthermore, we will also try to extend the framework proposed to support more emerging OpenMP detection tools.

\bibliography{Hierarchical_new}

\appendix

\end{document}